\documentclass[extra,mreferee]{gji}
\usepackage{timet,color}
\usepackage{graphicx}
\usepackage[urlcolor=blue,citecolor=black,linkcolor=black]{hyperref}

\title[Seismic and acoustic signals from the 2014 `Interstellar Meteor']
  {Seismic and acoustic signals from the 2014 `Interstellar Meteor'}
\author[Fernando \textsl{et al}]
  {Benjamin Fernando$^1$, Pierrick Mialle$^2$, G\"{o}ram Ekstr\"{o}m$^3$, Constantinos Charalambous$^4$, \\
  Steven Desch$^5$, Alan Jackson$^6$, Eleanor K. Sansom$^{7,8}$ \\
  $^1$ Department of Earth and Planetary Sciences, Johns Hopkins University, Baltimore, Maryland, United States  \\
  $^2$ Comprehensive Nuclear-Test-Ban Treaty Organisation, Vienna International Centre, Vienna, Austria \\ 
  $^3$ Lamont-Doherty Earth Observatory, Columbia University, New York, United States \\
  $^4$ Department of Electrical and Electronic Engineering, Imperial College London, London, United Kingdom \\
  $^5$ School of Earth Science and Space Exploration, Arizona State University, Tempe, Arizona, United States \\
  $^6$ Department of Physics, Astronomy, and Geosciences, Towson University, Baltimore, Maryland, United States \\
  $^7$ International Centre for Radio Astronomy Research, Curtin University, Perth, Australia \\
  $^8$ Space Science and Technology Centre, School of Earth and Planetary Science, Curtin University, Perth, Australia
  }
\date{XX}
\pagerange{\pageref{firstpage}--\pageref{lastpage}}
\volume{XX}
\pubyear{XX}

\begin{document}

\label{firstpage}

\maketitle

\begin{summary}
 We conduct a thorough analysis of seismic and acoustic data from the so-called `Interstellar Meteor' which entered the Earth's atmosphere off the coast of Papua New Guinea on 2014-01-08. We conclude that both previously-reported seismic signals are spurious - one has characteristics suggesting a local vehicular-traffic based origin; whilst the other is statistically indistinguishable from the background noise. As such, previously-reported localisations based on this data are spurious. Analysis of acoustic data cannot provides a best fit location estimate which is very far ($\sim$170~km) from the reported fireball location. Accordingly, we conclude that material recovered from the seafloor and purported to be from the meteor is almost certainly unrelated to it, and is likely of more mundane (non-interstellar) origin.

\end{summary}


\section{Introduction}
\subsection{Interstellar objects in the Solar System}

The discovery of the first interstellar object (1I/‘Oumuamua) in 2017 demonstrated that small bodies of extrasolar origin occasionally transit through our Solar System \citep{meech2017brief}, though the rate at which this happens is poorly constrained at present \citep{lsst}. Once within the Solar System, some of these objects are likely captured into bound orbits \citep{napier2021capture}. Presumably (and less frequently), some occasionally collide with the Earth \citep{desch2022some}. Detecting such an `interstellar impact event' would be of enormous scientific value. 

Even remote observations of a small interstellar meteoroid burning in the atmosphere would enable measurements of its composition, strength, and trajectory to be made, as is the case for Solar System meteors (e.g. \citet{fireballtheory}). However, despite extensive searching, no unambiguous, concrete detections of interstellar meteoroids have been made \citep{possible_can_ims}. Recent work has highlighted the significant challenges associated with confirming a meteoroid's interstellar status \citep{hardtofind_ims}, as where interstellar trajectories are inferred that they are often the result of measurement uncertainties \citep{measurementerrors}. Recent work has also noted that these errors become proportionally larger as an object velocity increases, leading to greater uncertainty in trajectory determination for fast-moving meteors regardless of their origin \citep{worse_as_faster}. This means that any candidate interstellar meteoroid, which would be moving significantly faster than one originating within the Solar System, would come with more uncertainty in its measured trajectory. 

Interstellar objects large enough to survive passage through the Earth's atmosphere are likely much rarer than small ones which burn up in the atmosphere, assuming that interstellar object populations follow similar mass-frequency distributions to those from within the Solar System (e.g. \citet{meteor_powerlaw}). We would also expect that an interstellar meteoroid would be less likely to have fragments survive atmospheric passage to impact the surface than an equivalent Solar System meteoroid; given the greater atmospheric disruption expected from its higher speed. Nonetheless, recovery of fragments from an interstellar meteoroid would be of even greater scientific value than simply recording one burning up in the atmosphere, as it would enable precision analytical geochemistry techniques to be applied to extrasolar samples for the first time. 

Where fragments do reach the surface, these likely represent only a small fraction of the meteoroid's initial mass \citep{popova2019modelling}. Potential recovery of fallen material from a given meteor would be predicated on tracking it through the atmosphere and inferring likely fall locations from its trajectory, given that the impact of fragments on the ground itself is almost never recorded \citep{carancas}.

\subsection{Tracking meteors in the atmosphere}

The tracking of meteoroids through the Earth's atmosphere is well-established \citep{mccord1995detection,edwards2008seismic}. 

From an electromagnetic perspective, optical/near-optical wavelength tracking using cameras is commonplace \citep{opticaltracking}; as is radio-wavelength tracking using either passive FM reflections or active radar \citep{radiodetection,radartracking}. 

From a mechanical perspective, both infrasound (low-frequency sound) and seismic tracking of meteoroids are also in widespread use \citep{edwards2008seismic,edwards2009meteor,silber2018infrasound}. In the case of infrasound, meteor sonic booms (non-linear shockwaves) may be detected close to the source (either from supersonic atmospheric entry or eventual explosion). Further away, shockwaves decay into linear sound waves which may propagate for many thousands of kilometres. In the seismic case, meteoroid detections can involve recording infrasound waves on a near-surface seismometer through induced ground deformation (e.g. compliance), or coupling of acoustic energy into the sub-surface and its subsequent propagation through the ground (often as surface waves). These techniques have even been recently extended to seismic detections of meteoroids on Mars \citep{garcia2022newly,daubar2023two}.

For the largest bolides, seismic detections of air-to-ground coupled Rayleigh waves can be made thousands of kilometres from the source \citep{karakostas2018inversion}. For smaller bolides, seismic detections are generally restricted to hundreds of kilometres away \citep{neidhart2021statistical}; but detection of infrasound signals ($<$20~Hz) can still be made at great distances \citep{pilger2018large}. 

This can be attributed to the greater sensitivity of most infrasound sensors (which include mechanical filters to suppress wind noise and turbulence) to atmospheric pressure changes. In general, seismometers are buried and hence not directly coupled to the atmosphere but instead are more likely detect the complex deformation of the ground by the atmospheric pressure field in the vicinity of the seismometer \citep{sorrells1971preliminary}. This also results in modification of the recorded waveform on a seismometer by its passage from the air into the ground \citep{olivieri2023optical}. 

For some events, both infrasound and optical detections of the same meteoroid are made, enabling validation of shock models \citep{silber2015optical}; though evidence suggests that most optically detected meteors do not produce detected infrasound. Direct correlations between photometric light curves and seismic/infrasonic waveform features are not expected - not least because the meteor does not behave as a seismic point source (rather, a conical source with a hyperbolic footprint upon the ground); and because infrasonic emission is not constant along the meteor's trajectory but rather higher at lower altitudes \citep{silber2014optical}. This expectation is supported by observations of hypersonic sample return capsules, for which both photometric and seismic/acoustic measurements have been made simultaneously (e.g. \cite{yamamoto2011}). 

\section{The 2014 `Interstellar meteor'}

It was recently reported by Siraj \& Loeb (2022) that a meteor of interstellar origin entered the Earth's atmosphere off the coast of Papua New Guinea (PNG) on 2014-01-08. 

This identification is derived from an entry in the Centre for Near Earth Object Studies (CNEOS) fireball catalogue\footnote{\url{https://cneos.jpl.nasa.gov/fireballs/}}. This database records a fireball event occurring on 2014-01-08 at 17:05:34~UTC (2014-01-09 01:05:34~UTC+10/local). 

The fireball location is reported inconsistently: it appears in the CNEOS catalogue as being centred around 1.3$^o$S, 147.6$^o$E, and in the CNEOS bolide detection notification as centred around 1.2$^o$S, 147.1$^o$E. This translates to a difference of approximately 57~km once projected onto the Earth's surface. Based on CNEOS data, the fireball's angle of entry was approximately 31$^o$ with respect to the horizontal.

CNEOS reports the fireball as occurring at an altitude of 18.7~km, and consisting of three distinct peaks in the light curve\footnote{\url{https://cneos.jpl.nasa.gov/fireballs/lc/bolide.2014.008.170534.pdf}} (likely corresponding to three distinct disruption events) separated by two intervals of 0.1~s. The CNEOS reported total impact energy is relatively low, at 0.11~kt TNT$_e$. 

An unusually high velocity is reported in the catalogue of $(v_x, v_y, v_z)$ = (-3.4, -43.5, -10.3)~km/s in a geocentric Earth-fixed reference frame. This is indicative of an meteor travelling on a roughly NW-SE path with a speed of 44.8~km/s. As noted previously, measurement errors from optical tracking systems at these velocities are significant and small uncertainties can make an orbit which is actually bound within the Solar System appear unbound \citep{measurementerrors}. \citet{devillepoix2019observation} show that although energy estimates of CNEOS reports are very reliable, the location, airburst height, speed, and radiant are not. Crucially, this specific event was examined by \citep{worse_as_faster} who concluded that the peculiarly high velocity was most probably caused by a measurement error. 

\subsection{Seismic data from the `Interstellar Meteor'}

More recently, it has been reported by \citet{siraj2023} that an exact trajectory reconstruction for this meteoroid has been achieved through the use of seismic data. This reconstruction, based on a single station, claims to limit the fireball epicentre to a highly localised area, variously reported as being `16~km$^2$' in area \citep{siraj2023} or a `1 km-wide strip' \citep{loeblpsc}. 

We note that the reported precision of this measurement is far greater than comparable efforts reported using seismic methods in the literature - for example, the 2012 Sutter's Mill meteorite entry was recorded by eight seismometers tens of kilometres away, but the effective point-source altitude inferred from this data had a $\pm$10.9~km altitude uncertainty \citep{jenniskens2012radar}, corresponding to a surface-projected areal uncertainty much larger than 1~km$^2$. Similarly, the 2000 Mork\'{a}va meteorite's altitude of acoustic emission could only be confidently constrained to `$\sim$30-40~km' \citep{brown2003moravka}, despite detection by more than a dozen seismic stations within 180 km of the ground track (with closest $\sim 1$~km). These broad uncertainties derive from the substantial impact that atmospheric conditions can have on sound propagation \citep{silber2014optical}, the modification of the wavefield by coupling into the ground \citep{olivieri2023optical}, and the challenges of accounting for the extended nature of a hypersonic source (e.g. \citep{henneton2015numerical}). 

Nonetheless, the reported seismic-derived localisation for the candidate 2014 `interstellar meteor' was then used to determine the search area for an oceanographic expedition \citep{loeb2022overview}. This expedition, which sampled a reported area of 0.06~km$^2$ (equivalent to a 250 x 250~m square) reported the recovery of fallen material from the bolide; identified as metallic spherules of extrasolar \citep{loeb2023discovery} or even extra-terrestrial origin on the basis of their composition \citep{loeb2024recovery}. We note that that this is by no means the consensus conclusion and has been heavily disputed, particularly in relation to the compositional analysis of the spherules \citep{desch2023critique,gallardo2023anthropogenic}. 

\subsection{Paper aims}

Regardless of the meteor's actual origin (interstellar versus interplanetary), we set out to examine the seismic data used in the fireball localisation. Our aim is to determine whether or not the derived location was (1) reasonable and (2) stated to an appropriate degree of precision; and if not, what the implications of this would be for the origins of the material recovered from the seafloor. We also evaluate whether acoustic data not considered previously can be used to derive a location, and if so, whether it is compatible with a seismically-derived one. 

\section{Methodology}

\subsection{Seismic and infrasonic data}
\label{sec:seismicdatasource}

In this study, we use data from seismometers on Manus Island, Papua New Guinea and Coen, Queensland, Australia (as per \citet{siraj2023}). 

\begin{figure*}
    \includegraphics[width=\textwidth]{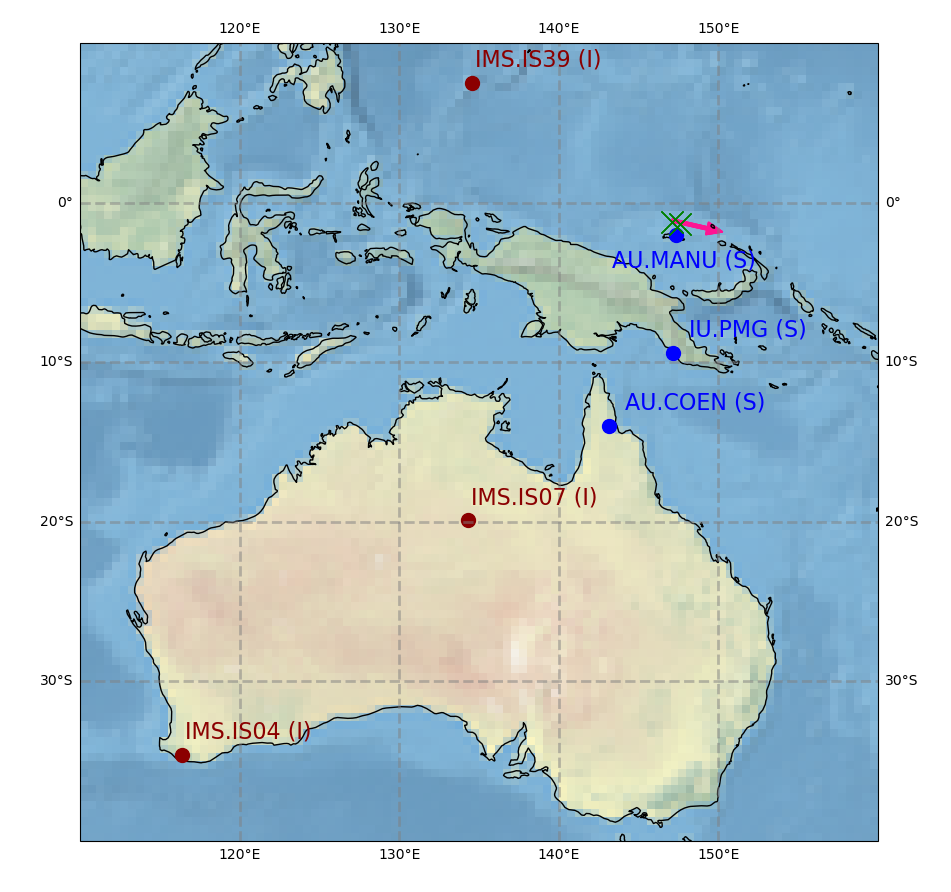}
     \caption{Map of seismic (blue) and infrasound (brown) stations used in this study. The two reported bolide locations are shown as green crosses (CNEOS bolide detection notification location is westernmost, the CNEOS catalogue location is easternmost\textit{}). The approximate trajectory of the meteor is shown as a pink arrow; note that the arrow's length is illustrative only and is not representative of travel distance or any other physical quantity because only an instantaneous meteor velocity is reported by CNEOS. Figure produced using Cartopy.}
     \label{fig:map}
\end{figure*}

Additionally, we consider seismic data from the seismic station at Port Moresby, Papua New Guinea; as well as infrasound data from the Comprehensive Test Ban Treaty Organisation's International Monitoring System (IMS) stations at Shannon (Western Australia), Warramunga (Australian Northern Territory), and Babeldaob Island, Palau. IMS stations are designed to detect signatures of nuclear tests \citep{okal2001t}, but are also sensitive to meteor-generated infrasound at distances of hundreds to tens of thousands of kilometers, depending on source location and yield \citep{pilger2018large}. The locations and codes of these stations are summarised in Table \ref{tab:stations} and shown in Fig \ref{fig:map}. 

\begin{table*}
    \centering
    \begin{tabular}{cccccc}
        \textbf{Name} & \textbf{Location}  & \textbf{Latitude}  & \textbf{Longitude} & \textbf{Distance (km)} & \textbf{Bearing (degrees)} \\
        AU.MANU (S) & Manus Island, PNG & -2.0432 & 147.3662 & 86 & 198\\
        IU.PMG (S) & Port Moresby, PNG & -9.4047 &	147.1597 & 897 & 183 \\
        AU.COEN (S) & Coen, QLD & -13.9674 & 143.1749 & 1483 & 199 \\
        IMS.IS39 (I) & Palau & 7.5 & 134.5 & 1748 & 304 \\
        IMS.IS07 (I) & Warramunga, NT & -19.9 & 134.3 & 2516 & 214 \\
        IMS.IS04 (I) & Shannon, WA & -34.6 & 116.4 & 4907 & 218 \\

    \end{tabular}
    \caption{Table 1: locations of seismic (S) and infrasound (I) stations used in this study. The AU network code indicates stations operated by Geoscience Australia's Passive Seismic Monitoring Network, IU indicates those of the Global Seismograph Network. IMS denotes stations operated by the Comprehensive Test Ban Treaty Organisation's International Monitoring System. Note that nearby seismic (IU.RABL) and IMS (IMS.IS40) stations not included in the above list were either offline on the day in question, or were not yet installed. Distances given are the great circle surface distances on the WGS84 ellipsoid including flattening of 0.0033528. Bearings toward sensors are given relative to grid north. Both assume a fireball position at the CNEOS catalogue location (1.3$^o$S, 147.6$^o$E).}
    \label{tab:stations}
\end{table*}

Note that the much greater distance at which meteor-generated signals are detectable via infrasound as opposed to via seismics means that we consider IMS stations further from the source than their seismic counterparts. Exact detection thresholds are highly situation-dependent, but data from the Australian Fireball Network relevant to this area suggests that bolides of this size are not detectable via seismic methods from more than 200~km away, but may be detected on infrasound upward of 1000~km away \citep{neidhart2021statistical,pilger2018large}. 

\subsection{Seismoacoustic setting}

In order to contextualise this event, we briefly describe the seismic setting of the south-west Pacific Ocean. This is a region which is still sparsely instrumented, but extremely tectonically \citep{denham1969distribution} and volcanically \citep{johnson1988volcanism} active. As such, significant seismicity of non-impact origin is expected. 

Furthermore, we note that many of the seismic stations (e.g. AU.MANU) are expected to be heavily contaminated by `cultural' noise, being close to human habitation including roads and industrial buildings. IMS infrasound stations are arrays of individual stations, typically composed of 4 to 8 sites each equipped with a wind noise reduction system. These systems assist in reducing incoherent noise created by local wind and atmospheric turbulence. As far as possible, IMS stations are installed in secluded environments, and thus are expected to be less affected by anthropogenic noise than seismic stations. 

In addition, environmental noise contamination at the seismic stations is expected to be significant, as seismometers located on islands in the Pacific are expected to experience high levels of ocean-generated primary and secondary microseismic noise \citep{webb1998broadband}. Additionally, this event occurred in January, which is the time of the southwest monsoon, bringing with it more inclement weather and stronger seismic noise associated therein \citep{tanaka1994onset}.

\subsection{Data Processing}

Instrument responses are removed for the seismic stations through deconvolution via ObsPy \citep{beyreuther2010obspy}. Unless otherwise stated, data are order-8 (order-4 with zerophase correction) Butterworth-bandpass filtered between 100~s and 15~Hz. We note that the described methods \cite{siraj2023} do not appear to include these key steps.

\section{Results and discussion}

\subsection{Seismic data from station AU.COEN}

\begin{figure*}
    \includegraphics[width=\textwidth]{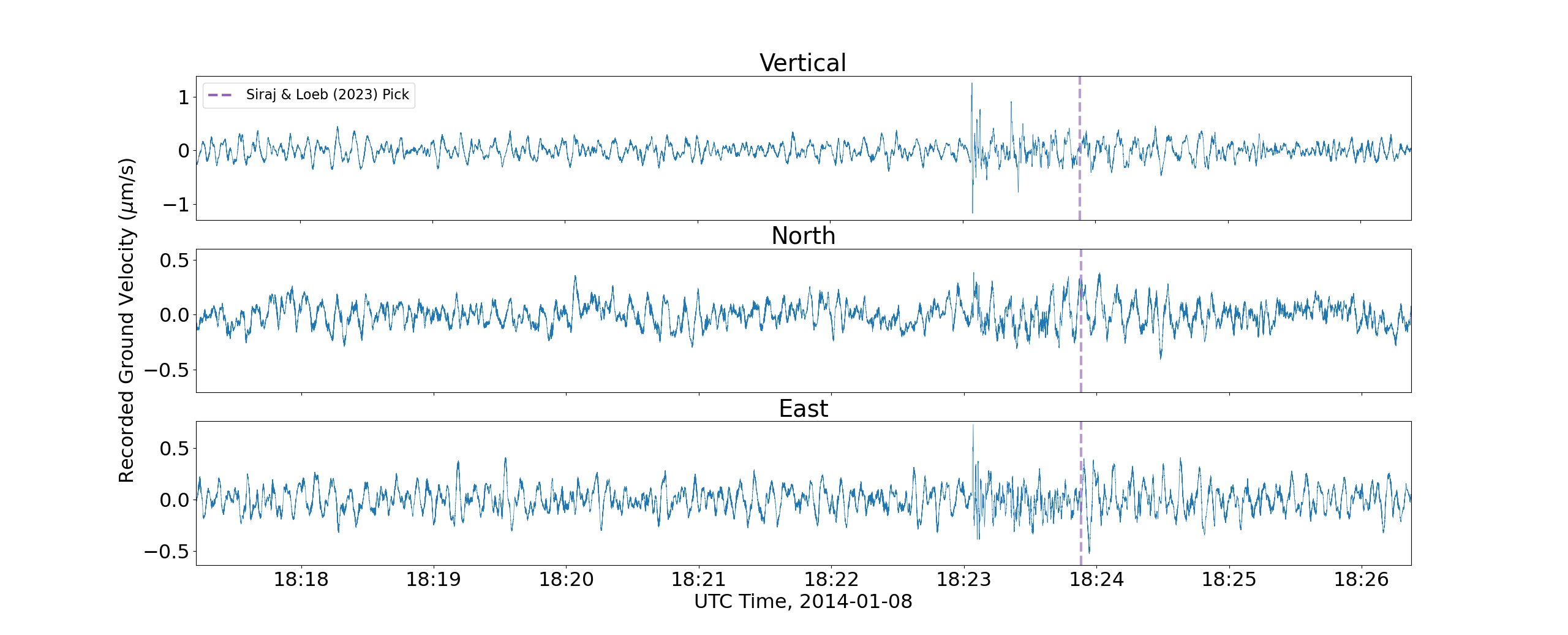}
     \caption{Data from the AU.COEN seismic station. The arrival time as picked by Siraj \& Loeb (2023) is shown as a dashed vertical purple line.}
     \label{fig:coen}
\end{figure*}

We begin by analysing the data from the broadband seismometer located at Coen, Queensland - a distance of 1483~km from the epicentre in the CNEOS catalogue. Siraj \& Loeb (2023) report an infrasound arrival at 18:23:53~UTC, some 78~minutes after the fireball. This data is shown in Fig. \ref{fig:coen}. 

As per Fig. \ref{fig:coen}, no seismic arrival can be seen at or immediately around 18:23:53~UTC. The arrival of some energy can be seen at 18:23:03~UTC, around 50~s earlier. This signal can clearly be excluded as being infrasound generated by the bolide for the following reasons: 

\begin{itemize}
\item For an object disintegrating at an altitude of 18.7~km, the line-of-sight sound horizon is approximately 500~km. As such, direct propagation (i.e. without reflection or refraction) of acoustic waves is impossible. Whilst waveguiding in the atmosphere or reflection from the troposphere are possible and can channel acoustic waves over long distances \citep{edwards2009meteor}, they would increase propagation lengths and travel times considerably \citep{silber2018infrasound}; rather than producing an early arrival.
\item Regardless, the path from the fireball to the AU.COEN station passes over areas of very steep topography (in excess of 4000~m over the New Guinea Highlands). This would be expected to disrupt both infrasound propagation and atmospheric waveguides themselves \citep{mckenna2012topographic}.
\item When rotated into a source-centred coordinate system (with an origin assumed at the CNEOS-catalogue fireball location), the arrival at AU.COEN shows a strong out-of-plane (transverse) signature. This is the opposite of what would be expected from either atmospheric waveguiding or sound waves reflecting off the ocean's surface - in both these cases, out-of-plane scattering is expected to be relatively weak. 
\item Empirical upper limits for the seismic detection of fireball signatures in Australia have been proposed \citep{neidhart2021statistical}. Using the same seismic network (though tested on smaller events, up to 0.005kt TNT$_e$), a source-receiver distance limit of 200~km is proposed. Although this event is around 20 times larger, it is nearly 8 times further away and hence and a spherical $\frac{1}{r^2}$ spreading would not be expected to bring its signal above the detectability threshold at AU.COEN. Whilst transmission losses can be reduced by assuming a waveguiding action, again this is not compatible with the measured travel times or proposed propagation geometries of \citet{siraj2023} as noted in the points above.
\end{itemize}

For these reasons, we conclude that no signal associated with the supposed interstellar meteor is detected at AU.COEN. 

\subsection{Data from station IU.PMG}

Next, we consider data from station IU.PMG, located on a similar azimuth (183$^{o}$) to AU.COEN and AU.MANU (199$^{o}$ and 198$^{o}$ respectively). This station was not examined by \citet{siraj2023}. These plots are shown in the supplement to this paper. In short, no appreciable seismic signal of any kind is apparent above the noise floor of this station between the purported arrival times at AU.MANU and AU.COEN. This is further evidence that the signal at AU.COEN is spurious; we would not in general expect an infrasound signal to be missing on a nearer seismometer, but be present on a more distant seismometer at the same azimuth and with a very similar noise floor (on the order of $\sim$0.5$\mu$m/s). 

\subsection{Seismic data from station AU.MANU}

Next, we consider seismic data from station AU.MANU, located on Manus Island, Papua New Guinea. This station is much closer to the reported fireball location ($\sim$86~km), and well within the distance to which infrasound could theoretically propagate.

\begin{figure*}
    \includegraphics[width=\textwidth]{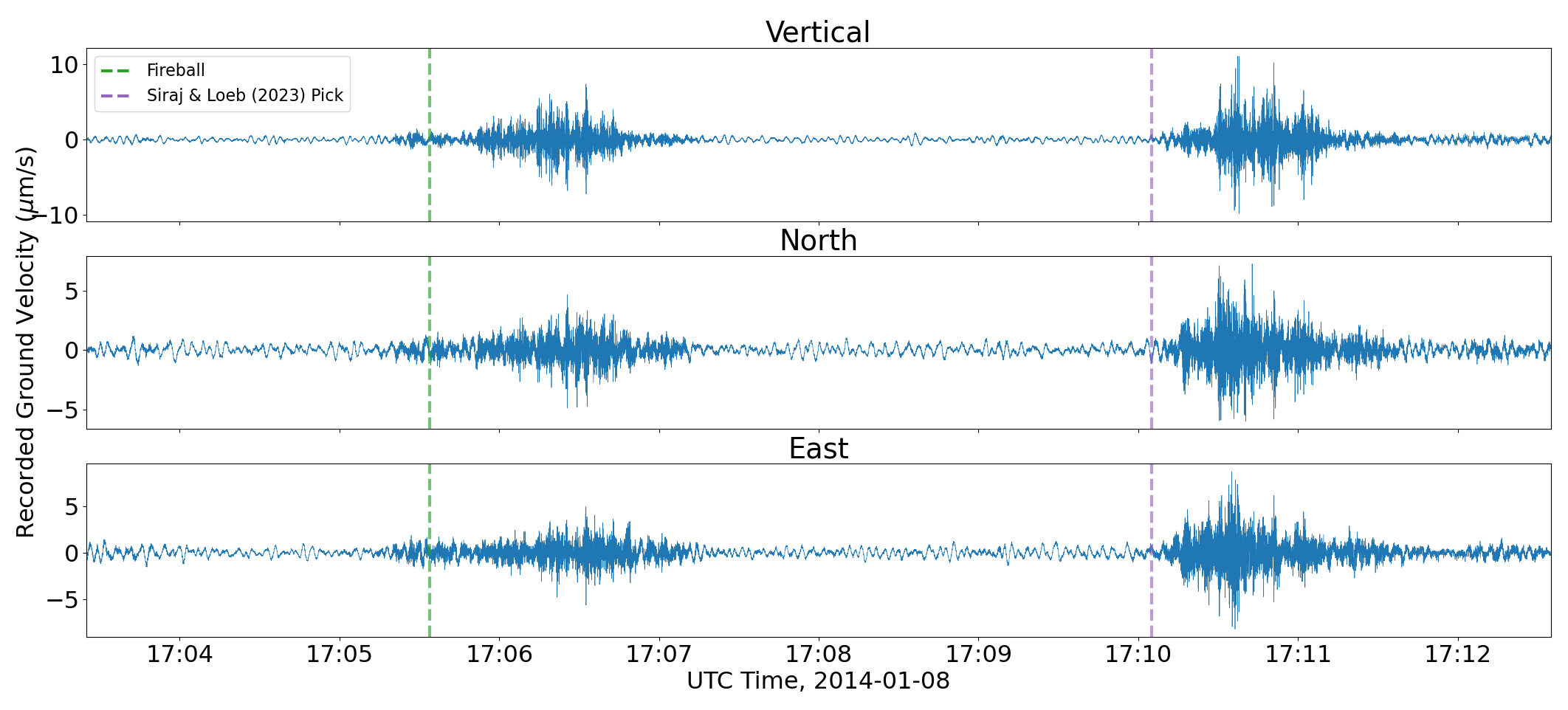}
    \caption{Data from the AU.MANU station. The arrival time as picked by Siraj \& Loeb (2023) is shown as a dashed vertical purple line, whilst the fireball origin time is marked with a dashed green vertical line.}
     \label{fig:manu}
\end{figure*}

Seismic data from this station are shown in Fig. \ref{fig:manu}. The event picked as meteor-generated infrasound begins at 17:10~UTC. The reader may wish to access the CNEOS lightcurve here\footnote{\url{https://cneos.jpl.nasa.gov/fireballs/lc/}}. 

We note that the light curve has a distinct three-peaked structure, which Siraj \& Loeb identify in the above seismogram as well (and use to derive a distance constraint). They suggest that this is due to some of the meteor infrasound reflecting off the ocean's surface. Whilst bolides have been detected in this manner \cite{pilger2020}, this technique is generally only possible for the largest bolides (e.g. the 173kT Bering Sea bolide of 2018, which had a yield a thousand times larger than this event). 

We do not consider such identifications in the seismogram as robust, as they are only weakly apparent in the vertical component and totally absent in the horizontals. Anyhow, the longer average source-reciever propagation distance over the ocean, coupled with lower signal-to-noise ratios from ocean-land interactions, generally make source identification and localisation very challenging for such events.

 At short distances, meteor-generated infrasound and seismic signals are generally short (seconds in length), and impulsive, with clear onsets. Exact waveforms can vary between classical N-waves (rapid overpressure followed by rapid underpressure, characteristic of sonic booms, \citep{edwards2009meteor}, or rapid onsets followed by slightly longer duration decays (e.g. \citep{kumar2017meteor}). This is distinctly different from the pattern observed in this event, which is extended and is nearly a minute in length. We note that longer-distance propagation (involving waveguiding or reflections off upper layers in the atmosphere) can produce more extended signals and `pulses' in meteor-generated infrasound; but again this would lengthen the travel time so as to no longer be commensurate with the estimates of Siraj \& Loeb (2023). 

\subsection{Seismic signal context}

\begin{figure*}
    \includegraphics[width=0.8\textwidth]{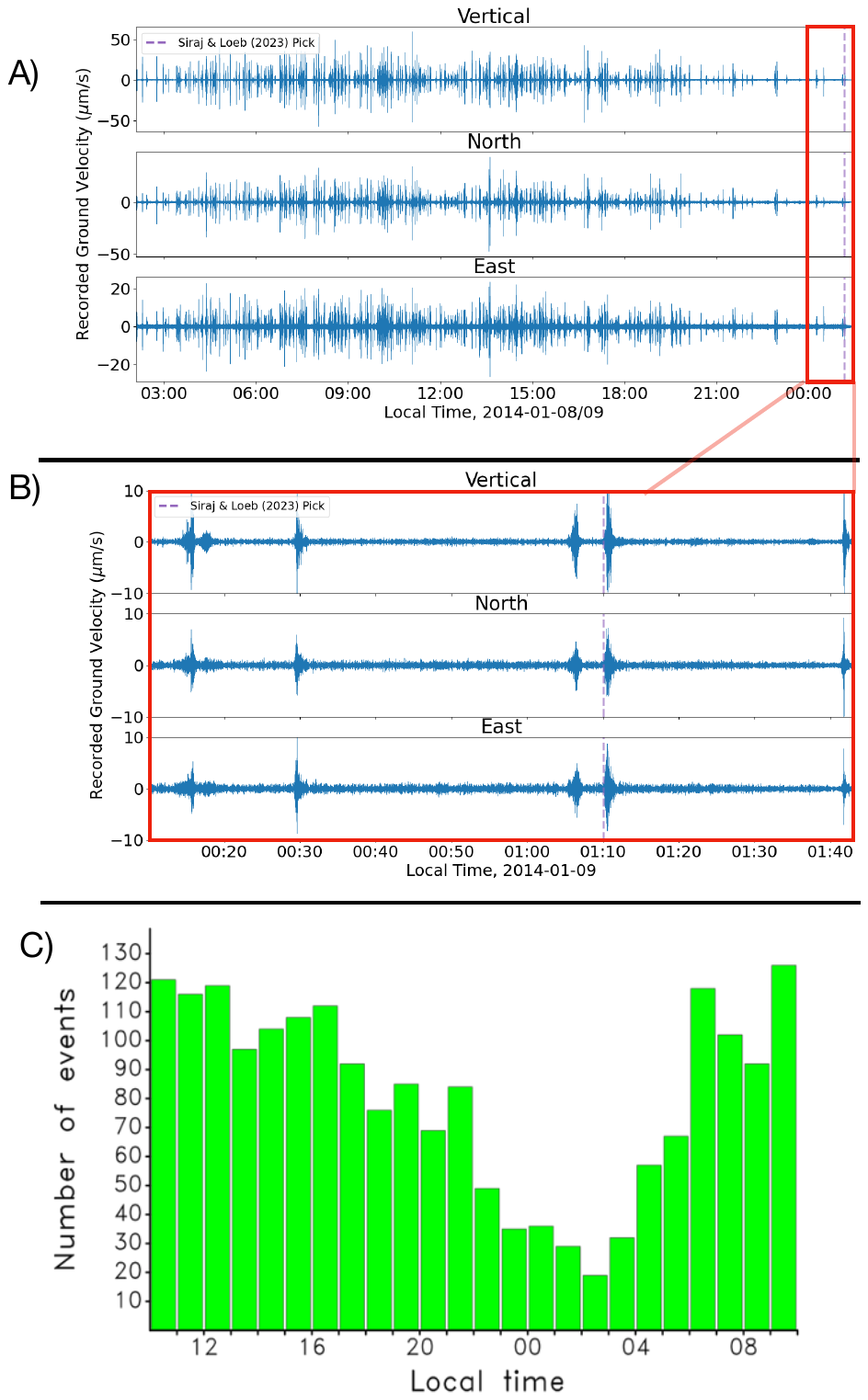}
    \caption{\textbf{A)} and \textbf{B)}: data from the AU.MANU station (axes now shown in local time) \textbf{A)} shows data for a 24~hour period before the event and 40~minutes after the purported meteor arrival (marked with a dashed purple line). \textbf{B)} shows a shorter context, with data from 60~minutes beforehand and 30~minutes after the end of the signal. A number of very similar signals (relatively rapid onsets, followed by slower decays) are apparent. \textbf{C)} shows the recorded number of `similar' seismic signals as a function of local time during the first two weeks of January 2014.}
     \label{fig:combined}
\end{figure*}

In order to contextualise the signal shown above, we consider the appearance of similar waveforms at AU.MANU seismic station around the time of the purported `interstellar meteor' arrival. This data is shown in Fig. \ref{fig:combined}. 

Our detection algorithm for `similar' signals involves searching through seismic data from this station for high-frequency signals which match in length, frequency, and waveform shape. Specifically, the detector searches for those which are approximately 60~s in length. Then, in order to compare waveform shape, the square of the 5-10~Hz band-passed signal is analysed. We define `similar' events to contain 95\% of the power (integrated over a 120~s window) within the central 60~s of that window. This definition is designed to be narrow enough to exclude other sources (e.g. earthquake signals), but allowing for some variation in signal parameters whilst maintaining a requirement for a peaked event with slower emergence and decay. The distribution in time of these signals shown in the Supplement.

Firstly, we note from Fig. \ref{fig:combined} \textbf{A)} that the signal at 01:10~local time (UTC+10) is not unusual in amplitude compared to other signals recorded during the day, which are presumably not of meteor-related origin. We also observe a clear variation throughout the day in the number of such signals observed, with far more between the hours of 04:00~local and 21:00~local. 

Statistically, this trend is bourne out in longer-term analysis. Fig. \ref{fig:combined} \textbf{C)} shows the number of `similar' events recorded as a function of local time during the first two weeks of January 2014. A strong diurnal patterning is observed. This is highly indicative of cultural (anthropogenic) noise. We also note a potential variation with day of the week (fewer strong signals on weekends) would further strengthen this conclusion, this is discussed further in the supplement to this paper).

Fig. \ref{fig:combined} \textbf{B} shows the 01:10~local time signal, and four other signals which are very similar in waveform characteristics (duration, amplitude, frequency content, and coda) during a 100~minute window around the event. Because the signal in question is highly similar to others occurring at AU.MANU on a frequent basis, it is reasonable to suspect that it is of a similar origin to them.

\subsection{Signal polarisation}

In order to elucidate what the origin of all these signals might be, we now consider the polarisation of the signal in question.

If the signal does represent cultural contamination noise, it is almost certainly produced locally and at or very near the ground's surface; and is likely to involve some interaction with the Earth's surface itself (e.g. vehicles driving, machinery shaking buildings, etc). Cultural noise is therefore expected to produce a signal with polarisation characteristic of a near-surface Rayleigh wave (elliptical retrograde). This would not be expected for a direct infrasound arrival. Furthermore, a Rayleigh wave detection could also be used to determine the signal's origin azimuth, and any temporal variation therein. 

Polarisation estimates are produced by filtering data between 8 and 10~Hz, and applying the method of \citet{roberts1989real} modified for Rayleigh waves. A sliding 1~second window is used to estimate the arrival angle every 0.33~seconds, and the associated uncertainty.  

\begin{figure*}
    \includegraphics[width=\textwidth]{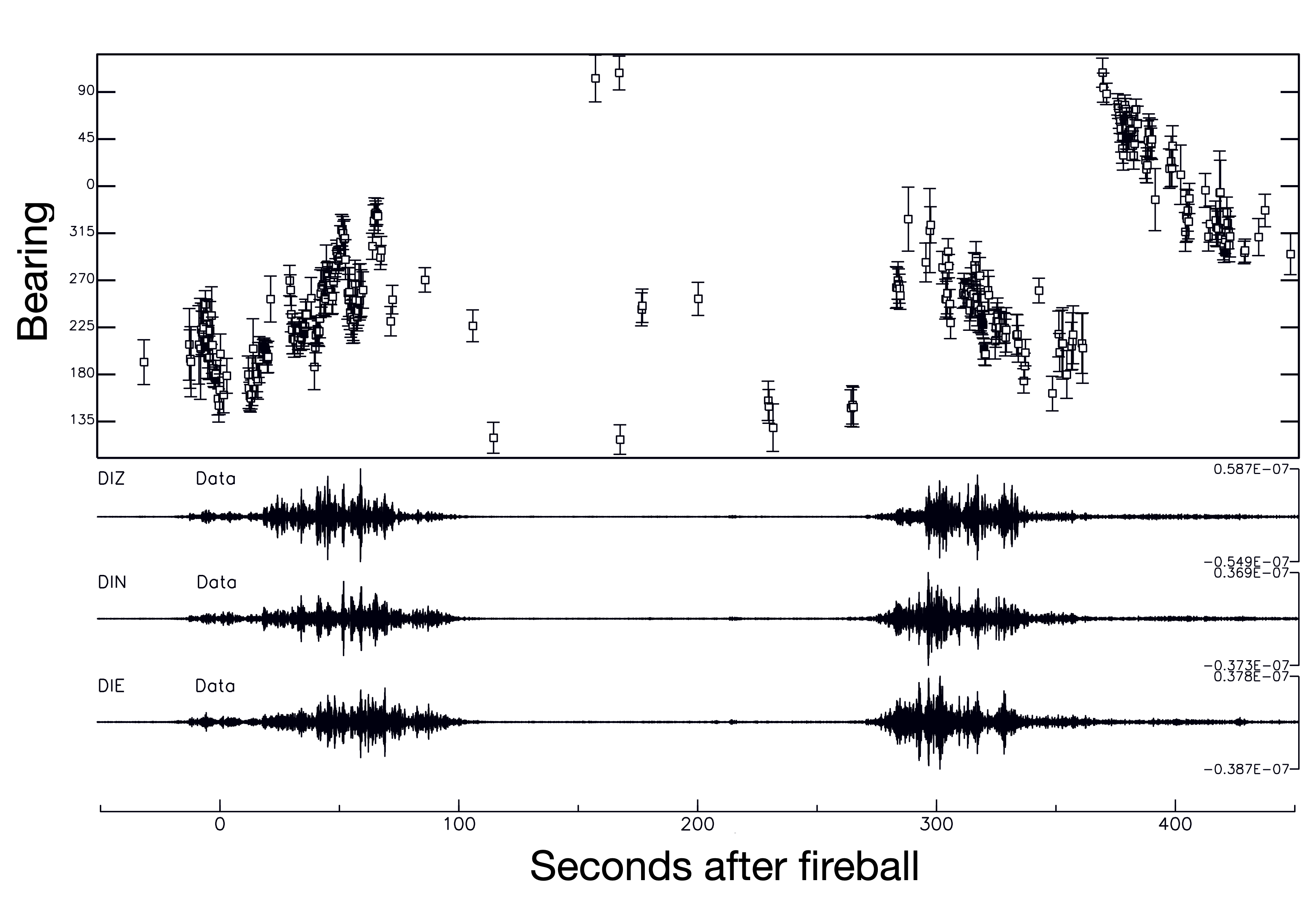}
    \caption{Signal polarisation for the purported meteoroid signal (beginning at $\sim$300~s, equivalent to 17:10~UTC) and a similar signal a few minutes earlier. Variation in azimuth over time is seen for both signals. Times are relative to the fireball origin, azimuths are measured clockwise from north as standard.}
     \label{fig:polarisation}
\end{figure*}

We find that the purported meteor signal has a clear Rayleigh wave polarisation, as does the signal occurring a few minutes beforehand. The variation of signal polarisation with time  is shown in Fig. \ref{fig:polarisation}, sampled every 1~s.

The first signal (01:05~local time) is emergent from a south-westerly direction, trending slowly northward over a period of around 100~s before dying away at a northward azimuth. Around 180~s later, the second signal (the purported meteor, 01:10~UTC) resumes at the same northward azimuth that the first signal ended at, and then trends back southward over another 100~s. The second signal has a longer coda, dying away on the same south-westerly azimuth to the origin of the first signal.

\subsection{Signal origin}

\begin{figure}
    \includegraphics[width=\textwidth]{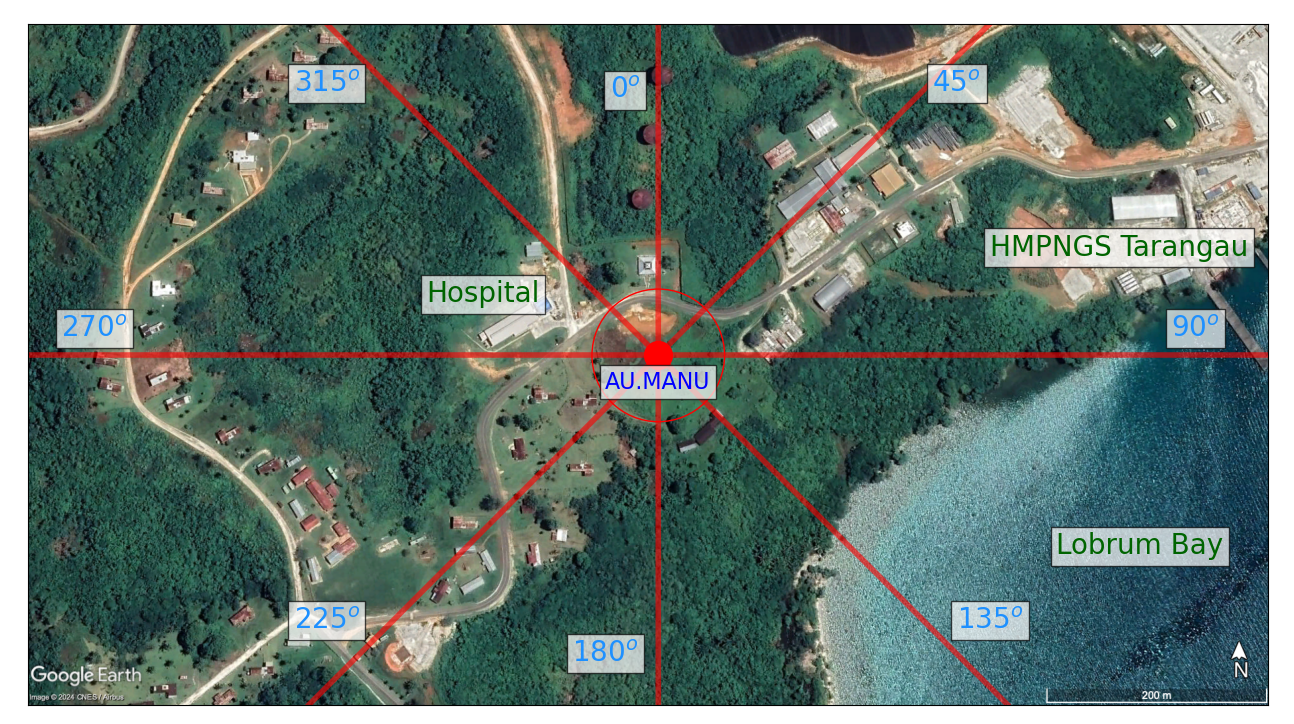}
    \caption{The local environment of the AU.MANU seismometer. North is upward and all bearings marked are relative to north. His Majesty's Papua New Guinea Ship Tarangau (a stone frigate/naval base also known as Lombrum Naval Base) is shown at upper right. A SW-N trending signal is consistent with passage along the road whose midpoint is bisected by the 225$^{o}$ bearing line. Data derived from CNES/Airbus imagery via Google Earth. Note that we choose a background image from 2024 rather than 2013 (the nearest image in time to the event in question), as the latter is lower-resolution and significantly obscured by cloud. We confirm nonetheless that none of the salient features labelled in the figure have moved.}
    \label{fig:manumap}
\end{figure}

Because the signal shows such a drastic rotation in azimuth over time ($\sim$90$^{o}$/minute), we can clearly exclude a teleseismic source - only a local source would be able to produce such a rapid change in the signal's direction of origin. For comparison to the polarisation plot in Fig. \ref{fig:polarisation}, we show a map of the AU.MANU locale in Fig. \ref{fig:manumap}. 

Noting the transition of the first signal from a south-westerly azimuth to a northward one, we see this is along the direction of the road whose midpoint is bisected by the 225$^{o}$ line. Similarly, the northward to south-westerly trend of the second (purported meteor) signal is consistent with passage along the road in the opposite direction, away from the naval base.

The Rayleigh wave signature of the AU.MANU waveforms and its rapid change in azimuth are highly suggestive of a vehicular (road traffic) source \citep{meng2021analysis}. The commonness and strength of such signals at AU.MANU is expected, given the roughness of roads (generating greater seismic noise) and the presence nearby of a military base. 

Specifically, the signal polarisation is indicative of a vehicle travelling up the road toward the seismometer, pausing in the vicinity of the hospital, before heading back in the reverse direction a few minutes later. We suggest that a simple and logical explanation for this behaviour would be a pick-up/drop-off of someone/something at the hospital. Once this signal is accounted for, no other unusual signal possibly associated with the meteor is observed in the data. 

Therefore, we simply conclude that the purported interstellar meteor was not recorded seismically; as is the case for the vast majority ($\sim$99\%) of optically-tracked meteors which are not even detected on more sensitive infrasound instruments \citep{silber2014optical}.

\subsection{Acoustic/Infrasound data}

Having demonstrated that no seismic arrival from the meteor is apparent at any of the three seismic stations we have considered, we evaluate whether a location estimate can be derived using infrasound data. As described in Sec. \ref{sec:seismicdatasource}, we make use of data from the International Monitoring System of the Comprehensive Nuclear-Test-Ban Treaty Organisation (CTBTO). The sensors (microbaromters) are expected to be sensitive to meteor-generated signals at much greater ranges than seismic networks (up to thousands or tens of thousands of kilometres for the largest events, \citep{pilger2018large, mialle2019}). 

\begin{figure}
    \includegraphics[width=\textwidth]{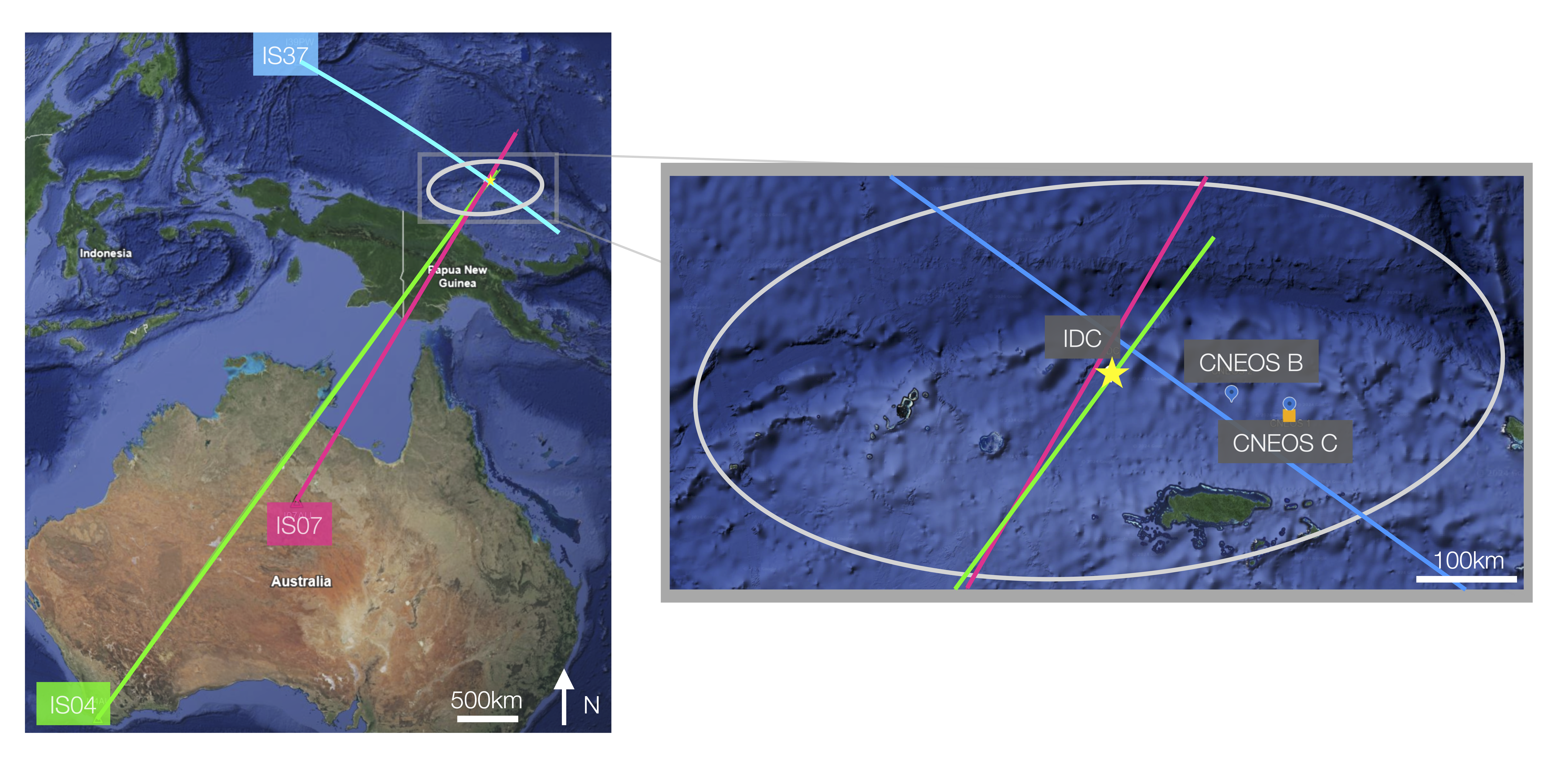}
    \caption{Derived back-azimuths from the three IMS stations which recorded the meteor's atmospheric entry. There is no single point of intersection between the three derived back-azimuths. Rather, a large `cocked hat' triangular shape is produced by the three individual intersections, as highlighted by the inset. The best-fit location from these measurements is highlighted with a yellow star. The box derived by Siraj \& Loeb (2023) from seismic data is shown as a orange square, the CNEOS-catalogue reported fireball location is marked with a blue pin labelled `CNEOS C', whilst the bolide notification reported fireball location is marked `CNEOS B'. The gray ellipse represents a 90\% confidence interval for the acoustic emission location}
    \label{fig:ctbto}
\end{figure}

Three stations record signals consistent with the atmospheric entry of a meteor at the time in question. Derived back-azimuths from these stations, as determined by the CTBTO's International Data Centre (IDC), are shown in Fig. \ref{fig:ctbto}.

As per Fig. \ref{fig:ctbto}, no exact location can be derived for the meteor's location based on infrasound, though a best-fit estimate is denoted by a yellow star. The `cocked hat' produced by the intersections of the three derived backazimuths is NE-SW orientated (approximately perpendicular to the meteor's trajectory). The 90\% confidence uncertainty ellipse for the acoustic-derived origin has semi-minor and semi-major axes of 186 and 388~km respectively, and an enormous area of 227,000~km$^2$. The best-fit estimate is at the centre of this ellipse, at a significant distance $\sim$170~km from the CNEOS catalogue location, and that of \citet{siraj2023}.
The uncertainty associated with the recorded back-azimuth directions and times of arrival is small, however these can only be interpreted in the context of the atmospheric model which is poorly known, producing the large error ellipse. Further studies of source properties, or triangulation with more than three stations, is precluded by the very low signal-to-noise ratio of this event (consistent with the low reported yield of 0.11 kt TNT$_e$ in the CNEOS catalogue). 

\subsection{Implications for recovery of fallen material}

We now consider what the implications of the above results are for the recovery of potential fallen material from this meteor. As noted previously, identifying fragment fall locations even where detections are made on multiple seismic or infrasound stations is extremely challenging \citep{brown2003moravka, jenniskens2012radar}. It is entirely possible, and potentially likely, that no material from this event would survive to reach the surface in large enough volumes or masses to be practicably recovered \cite{desch2023critique}. Therefore, this discussion should be considered illustrative of the fact that the search methodology of \citet{loeb2024recovery} was likely flawed, rather than a prediction of exact fall locations. 

In our analysis, we derived a best-fit location derived from acoustic data which is around 170~km from the \citet{siraj2023} location. Although the uncertainty ellipse includes the CNEOS location, this is not surprsing given how large the ellipse is.

Furthermore, it is highly unlikely that the CNEOS location is accurate anyhow. The phenomenon of mis-location in the CNEOS catalogue has been described in detail by \citet{devillepoix2019observation}. Using other high-fidelity sources; and they find that fewer than 50\% of CNEOS report locations are consistent with other measurements to within the stated error. We are therefore significantly more confident in our best-fit location than that derived from CNEOS. 

In light of this, we consider whether a strewn field centred around the acoustic-derived origin might potentially reach all the way (170~km or more) to the search area of \citet{loeb2024recovery}. 

The behaviour of small fragments post-fragmentation events is complex, and strongly dependent on their exact size and the current atmospheric conditions. The velocity changes associated with fragmentation are small compared to the meteor's initial ballistic velocity, and smaller material tends to continue on said ballistic trajectory with large fragments at first \citep{jenniskens2022bolide}. However, small fragments are also more affected by atmospheric drag \citet{toth2015kovsice, moilanen2021determination} and hence decelerate faster and travel a shorter downrange distance before hitting the surface \citep{passey1980effects}. 

In this event, the meteor's entry angle relative to the horizontal of 31$^{\circ}$ is a significant factor. Based on a fragmentation altitude of 18.7~km, fragments continuing on a trajectory of 31$^{\circ}$ would not be expected to travel more than 30-40~km. As noted, smaller fragments would travel shorter distances than larger ones as well, making the probability of the strewn field containing millimetre-sized spherules extending 170~km downrange very slim. We also note that even well-documented falls, where large masses of material are recovered, do not have strewn fields extending close to 170~km in length \cite{russell2023recovery, limonta2021fragmentation}. This line of reasoning indicates that the material recovered from the seafloor is highly likely to be unrelated to the meteor. 

Alternately, we may consider the problem by recognising that the best-fit location derived from acoustic data is only the most likely probability within a wide uncertainty ellipse. At 227,000~km$^2$ in area, it is impossible to practically search this area. 

Even if the strewn field is on the order of many kilometres wide and many tens of kilometres long (hence up to hundreds of kilometres square in area), this areal extent is still a tiny fraction of the 227,000~km$^2$ area of the uncertainty ellipse.

Given that the \citet{loeb2024recovery} search area of 0.06~km$^2$ is 0.0002\% of the total ellipse, the probability of this search area intersecting the true strewn field area is exceedingly small. We do not consider that the CNEOS-derived locations can be practically used to meaningfully constrain the ellipse any further, given the wide uncertainties reported by \citet{devillepoix2019observation}. Therefore, the overlap between \citet{siraj2023}'s derived origin and the CNEOS location are more likely a result of \citet{siraj2023} incorrectly fitting a sound-speed model with the aim of restricting the CNEOS location box, rather than this being a physically meaningful constraint on the location. 

Again, this is highly suggestive of the recovered material being unrelated to the meteor, simply because the localisation is so poor that the chances of a localised oceanographic expedition sampling the correct area are extremely small.  

\section{Conclusions}

We have examined data from three seismic stations: AU.MANU, AU.COEN, and IU.PMG. AU.COEN and IU.PMG show no arrival in the relevant time window of any kind, only background seismic noise, as expected at these distances. 

At AU.MANU, an arrival is apparent at approximately the time stated by \citet{siraj2023}. However, this arrival shows none of the characteristics expected of infrasound from a nearby meteoroid (e.g. as compared to \citet{edwards2008seismic}). It is very closely related to other signals of similar characteristics around the same time. 

Its polarisation is indicative of a Rayleigh wave, and its rapid change in azimuth of origin indicates a source very close to the seismic station. The change in azimuth with time is consistent with a moving vehicle traversing the road next to the seismometer toward a local hospital, before heading away again. This is also commensurate with observations that signals of this nature are common at AU.MANU, and show the diurnal variation expected of anthropogenic/cultural noise. We also expect some contamination of the signal from tectonic-origin seismic energy, though this is likely small (see supplement). Therefore, we conclude that the seismic arrival in question does not in fact represent an infrasound arrival from the meteor in question. 

Infrasound data from the CTBTO's IMS stations IS39, IS04, and IS07 did detect infrasound likely associated with the meteor. The signal-to-noise ratio is too poor to enable source characterisation or exact localisation, but the 90\% confidence uncertainty ellipse as bounded by the three derived back-azimuths is large ($\sim$270,000~km$^{2}$) and distant ($\sim$170~km) from both the CNEOS reported fireball location and the reported localisation from \citet{siraj2023}.

As such, the chances that the \citet{loeb2024recovery} searched the correct area of the seafloor are extremely small, if there was even any material to be recovered. Based on this analysis, we conclude with a high degree of confidence that the purported fragments of the meteor recovered from the seafloor have nothing to do with the fireball, regardless of whether the parent meteor was interstellar or Solar System bound. The spherules found by \citet{loeb2024recovery} are therefore likely of more mundane origin, with their purported unusual chemical signatures caused by contamination of terrestrial pollutants or chemically alteration by exposure to seawater \citep{gallardo2023anthropogenic}.

\begin{acknowledgments}
The authors acknowledge the data provided freely by CNEOS, the Geoscience Australia Seismic Network, the Global Seismograph Network, and the International Seismological Centre. We also acknowledge the traditional owners of the land upon which many of these stations are located. 

BF is funded by the Johns Hopkins University Blaustein Fellowship in Earth and Planetary Science. 
\end{acknowledgments}

\begin{dataavailability}
Data from CNEOS can be downloaded at \url{https://cneos.jpl.nasa.gov/fireballs/}. Data from the Geoscience Australia Seismic Network can be downloaded via the Federation of Digital Seismograph Networks \url{https://www.fdsn.org/networks/detail/AU/}. The International Seismological Centre Event Bulletin can be found at \url{http://www.isc.ac.uk/iscbulletin/search/catalogue/}; the quake shown from the Kuril Islands constitutes event number 603942758. 

Data from the International Monitoring System (IMS) of the Comprehensive Test Ban Treaty Organisation (CTBTO) can be requested in raw format directly from the CTBTO via a zero-cost contract with vDEC \url{https://www.ctbto.org/resources/for-researchers-experts/vdec/request-for-data}, under the terms of data sharing we are unable to release this directly. The views expressed herein are those of the author(s) and do not necessarily reflect the views of the CTBTO Preparatory Commission.

\end{dataavailability}

\bibliographystyle{gji}
\bibliography{mybibfile.bib}

\bsp %

\label{lastpage}

\end{document}

\label{firstpage}

\maketitle

\section{Introduction}

This supplement contains discussion of topics ancilliary to the main manuscript, which are included for completeness. 

\section{Data from IU.PMG}

\begin{figure*}
    \includegraphics[width=\textwidth]{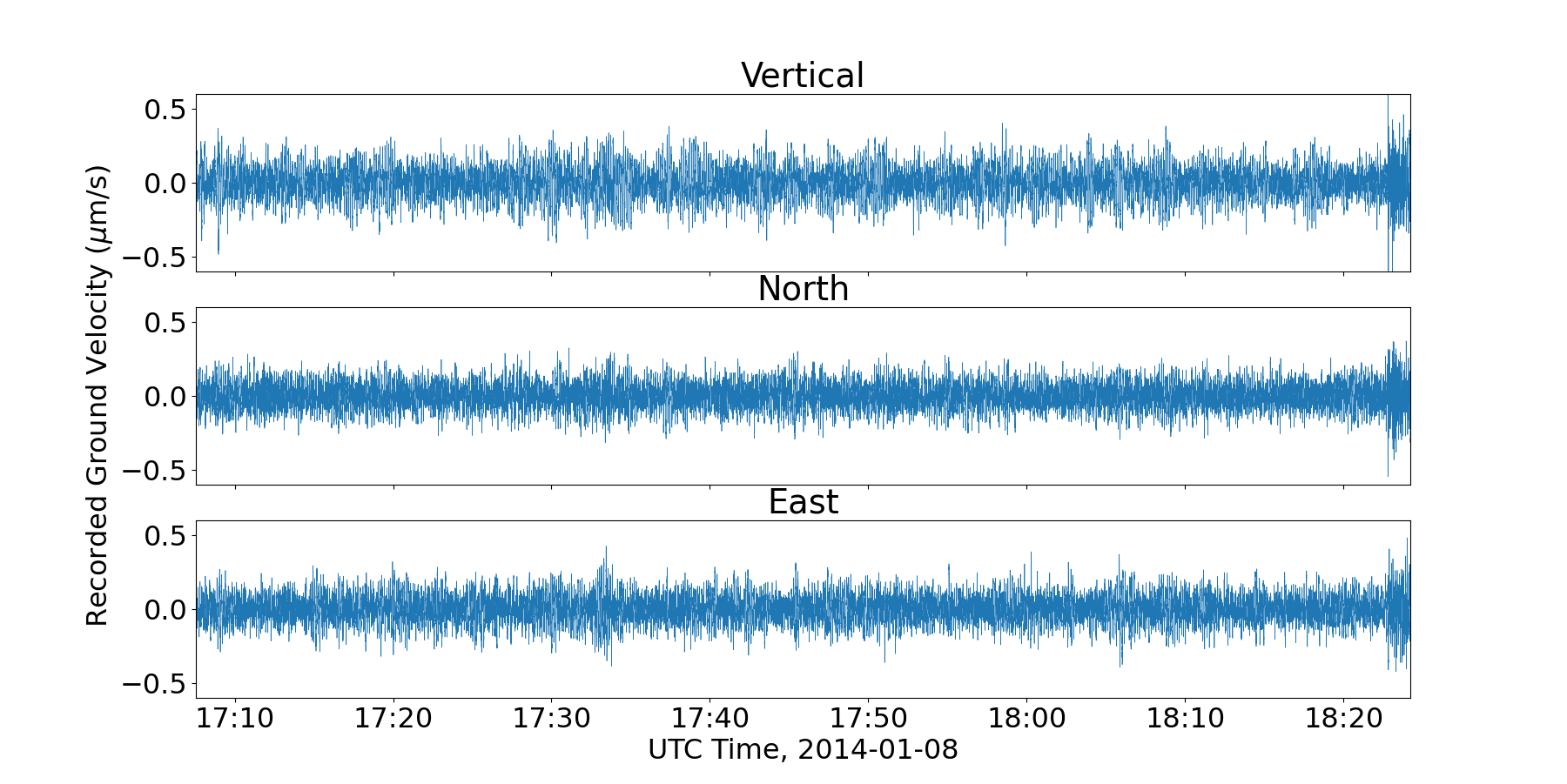}
     \caption{Data from station IU.PMG, for the period between the purported AU.MANU (17:10~UTC) and AU.COEN arrivals (18:23~UTC).}
     \label{fig:pmg}
\end{figure*}

Data from station IU.PMG, at Port Moresby, Papua New Guinea, are shown in Fig. \ref{fig:pmg}. No ballistic arrivals of any kind can be seen in the data. We consider it highly implausible that an infrasound signal would be detected at AU.MANU and AU.COEN, but not IU.PMG given that it is intermediate in distance between the two, on a very similar bearing from the reported fireball location, and has a similar noise floor to IU.PMG. We consider this as additional evidence that the AU.COEN signal is spurious. 

\section{Earthquake seismic signatures contaminating arrivals at AU.MANU}

For completeness, we also consider whether the period of time in question, around 17:10~UTC on 2014-01-08, may be contamined by earthquake-generated seismic arrivals as well as vehicular noise. 

To do this, we perform a search of the International Seismological Centre's Event database \cite{bondar2011improved}. As expected, this region is highly seismically active. For example, we note that an $M_w$ 4.7-5.3 earthquake with a predominantly reverse fault mechanism occurred in the Kuril Islands off Japan at approximately 16:50~UTC, around 15 minutes before the fireball and around 20 minutes before this detection (ISC Bulletin ID 603942758, hypocentre at 46.5$^{o}$N, 153.3$^o$E, 62~km depth). 

\begin{figure*}
    \includegraphics[width=\textwidth]{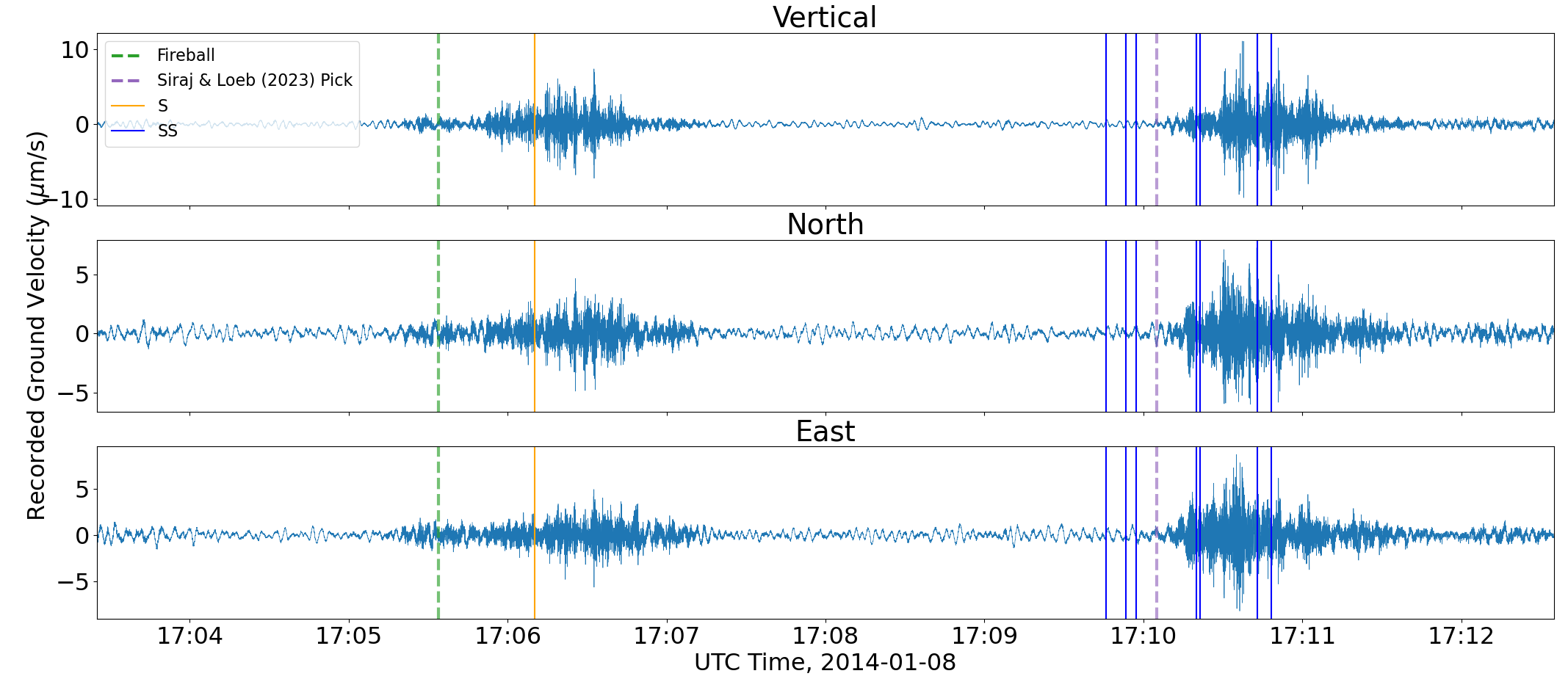}
     \caption{Seismic arrivals from ISC Event 603942758, overplotted on the signal recorded at AU.MANU. The fireball origin time and Siraj \& Loeb (2023) picks are denoted by green and red dashed lines, respectively. The orange solid line indicates TauP-calculated S wave arrival times, and blue solid lines are SS arrivals.}
     \label{fig:kuril_contam}
\end{figure*}

Calculation of expected travel times from this earthquake using the TauP raytracing method \cite{crotwell1999taup} indicates multiple seismic arrivals from the Kuril Islands quake are expected at AU.MANU in the window identified by \citet{siraj2023} as meteor-generated infrasound. These plots are shown in Fig. \ref{fig:kuril_contam}.

In the absence of full-waveform seismic modelling we make no quantiative estimate of how much these phases might be contributing to the seismic signal at this time; though the contribution is likely small given that most of the signal energy is above 5~Hz and mantle propagation at these distances heavily attenuates high-frequency ($>$1~Hz S-wave energy.

We remain confident in our overall conclusion that the seismic signal beginning at 17:10~UTC is dominantly a vehicle-generated Rayleigh wave; though caveat that some earthquake seismic energy may also be present. This is an additional reason that precision fitting of the seismic envelope as described by \citet{siraj2023} would not be valid without significant further modelling, even if this signal were infrasound from the meteor rather than a Rayleigh wave from a vehicle. 

\section{Similar signals at AU.MANU}

In Fig. 4 of the main paper we demonstrated that signals similar to that purported to be meteor-generated infrasound have a clear diurnal variation, which is indicative of cultural noise. The distribution of these signals in power and time is shown below in Fig. \ref{fig:weekend}. 

We note that there may be some evidence of a reduced density of the strongest signals (above the 10$^{-11}$ power level) during the weekend periods, if we exclude the period immediately after new year; however this analysis is not central to the paper and we note it only for completeness. 

\begin{figure*}
    \includegraphics[width=\textwidth]{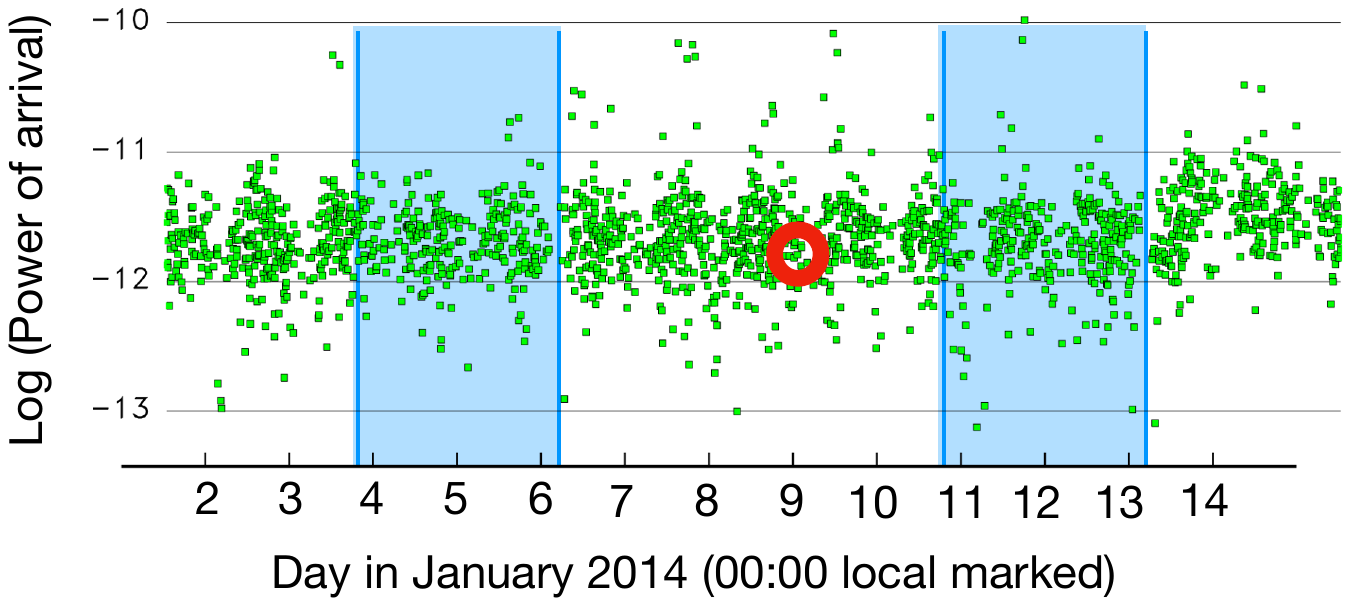}
     \caption{Similar seismic arrivals, plotted by incident velocity power against day. The red circle is centered on the purported meteor arrival. The two blue areas denote weekends (21:00~local Friday to 04:00~local Monday, commensurate with the `quiet period' in Fig. 4 from 21:00-04:00~local daily.}
     \label{fig:weekend}
\end{figure*}

\section{IMS Infrasound Data}

\begin{figure*}
    \includegraphics[width=\textwidth]{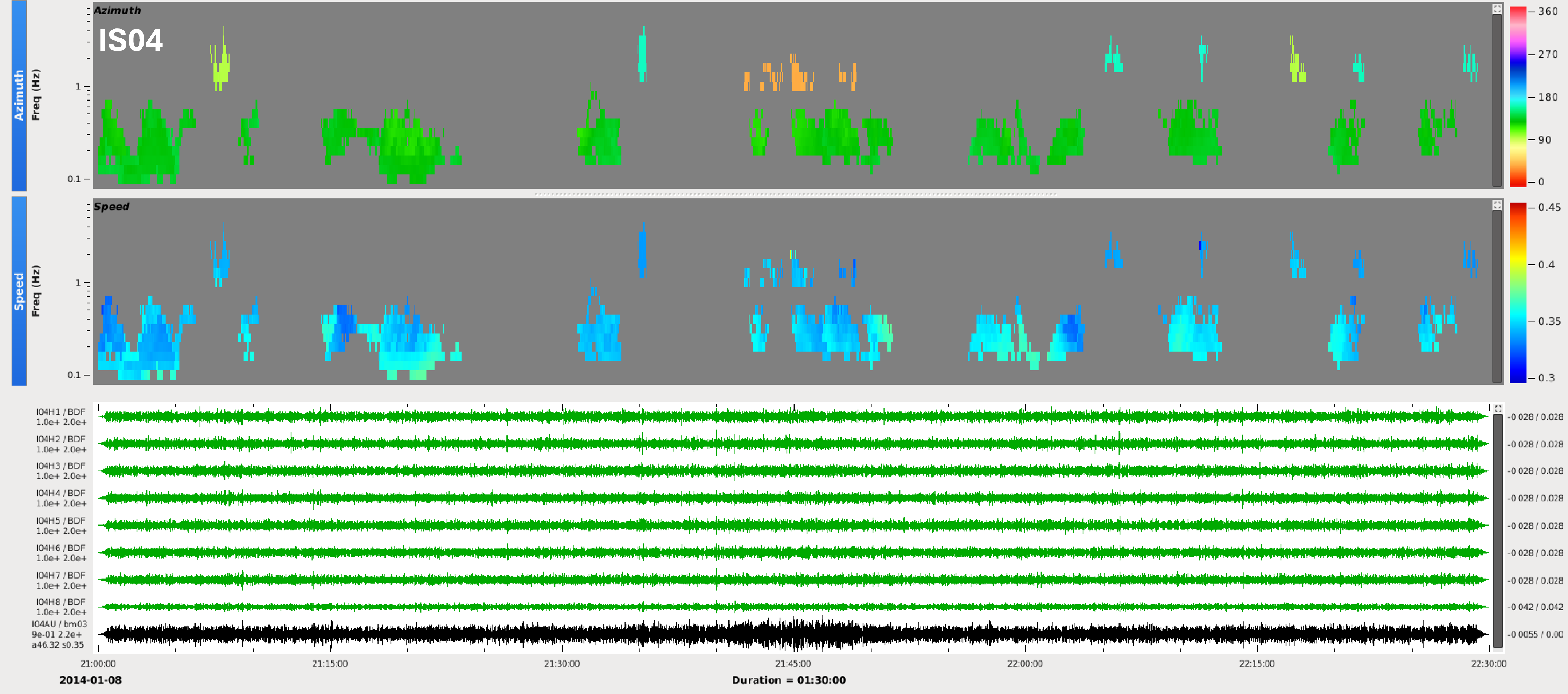}
     \caption{Infrasound data from station IS04 in Australia. Signal begins at around 21:41~UTC, in \textbf{orange} in the upper panel.\textbf{Top and Middle}  The top panels show back-azimuth and trace velocity estimates obtained via the PMCC method. The horizontal axis in each case is time, the vertical axis is a logarithmic frequency scale, and the colour bar indicates either the azimuth from north or speed in kilometres per second. \textbf{Bottom} The bottom panel contains the raw waveforms (green traces) and the beam (black trace) obtained using the estimated signal parameters. For this station, detected signals are in orange in the azimuth panel.}
     \label{fig:is04}
\end{figure*}

\begin{figure*}
    \includegraphics[width=\textwidth]{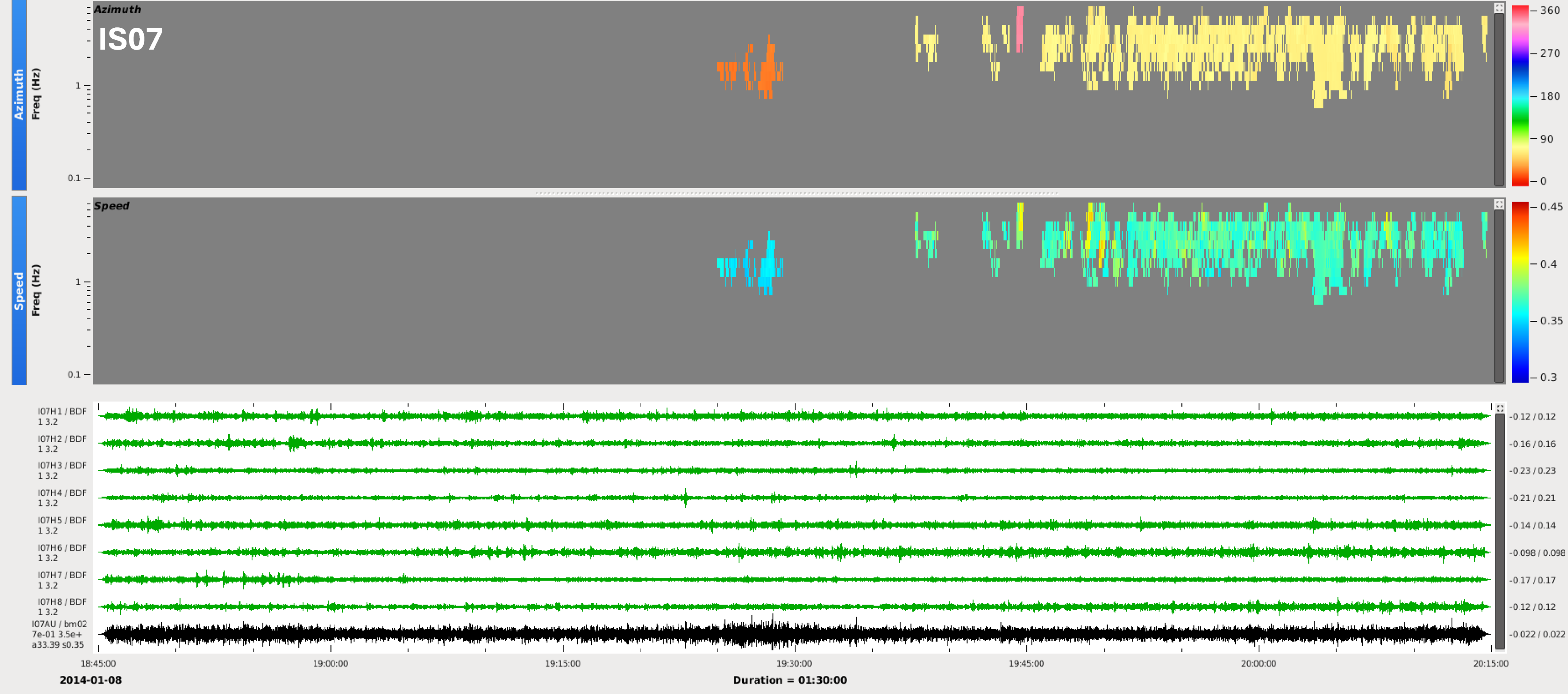}
     \caption{Infrasound data from station IS07 in Australia. Signal begins at around 19:25~UTC, in \textbf{orange} in the upper panel. \textbf{Top and Middle} The top panels show back-azimuth and trace velocity estimates obtained via the PMCC method. The horizontal axis in each case is time, the vertical axis is a logarithmic frequency scale, and the colour bar indicates either the azimuth from north or speed in kilometres per second. \textbf{Bottom} The bottom panel contains the raw waveforms (green traces) and the beam (black trace) obtained using the estimated signal parameters.}
     \label{fig:is07}
\end{figure*}

\begin{figure*}
    \includegraphics[width=\textwidth]{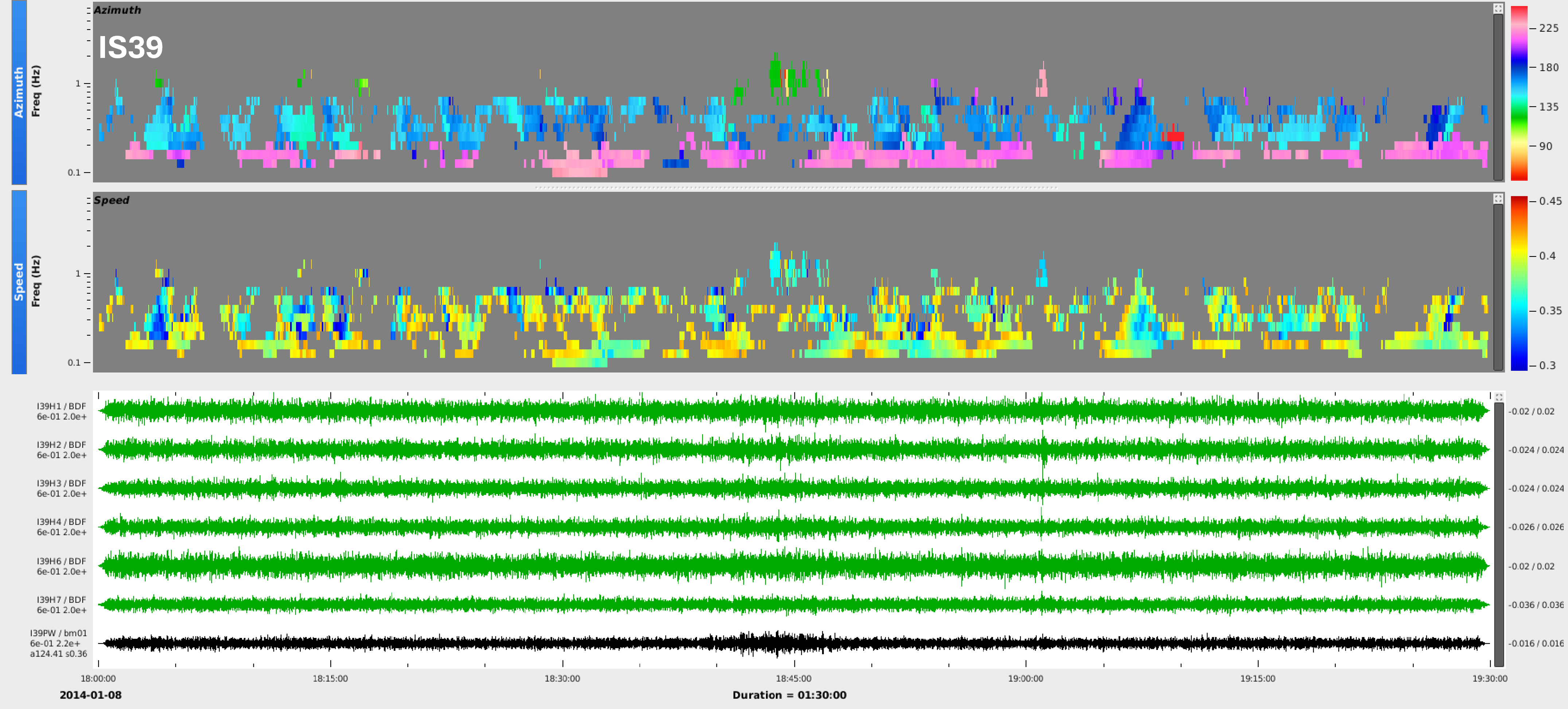}
     \caption{Infrasound data from station IS39 in Palau. Signal begins at around 18:40~UTC, in \textbf{green} in the upper panel. \textbf{Top and Middle} The top panels show back-azimuth and trace velocity estimates obtained via the PMCC method. The horizontal axis in each case is time, the vertical axis is a logarithmic frequency scale, and the colour bar indicates either the azimuth from north or speed in kilometres per second. \textbf{Bottom}The bottom panel contains the raw waveforms (green traces) and the beam (black trace) obtained using the estimated signal parameters.}
     \label{fig:is39}
\end{figure*}

Figures \ref{fig:is04}, \ref{fig:is07}, and \ref{fig:is39} show the infrasound data for the three IMS stations. The plots show the International Data Centre (IDC) of the CTBTO review interfaces for the infrasound data analysis. These plots make use of a multi-channel correlation method (DTK-GPMCC) processing. 

It should be noted that these signals all have weak signal to noise ratios (most apparent by the low elevation above the background noise of the beamformed signals, lowermost in each plot).

Figure captions include details of what features exactly constitute the identified signal - for IS04 and IS07 the signals are highlighted in orange (back-azimuths toward the north-east), and occur just above 1~Hz. They begin at 21:41 and 19:25~UTC respectively. At IS39 the signal is highlighted in green (back-azimuths toward the east) and begins at around 18:40~UTC and has a similar frequency range. 

Note that these colours are on the same scale between all three plots, the derived azimuths for IS04 and IS07 are similar because the stations are on roughly the bearing from the meteor location.

\bibliographystyle{gji}
\bibliography{mybibfile.bib}

\bsp %

\label{lastpage}